\documentstyle[twocolumn,aps]{revtex}
\begin {document}
%\draft

\wideabs{
\title{The density dependence of the transition temperature in a homogenous Bose fluid}
\author{J.D. Reppy, B.C. Crooker$^{a}$, B. Hebral$^{b}$, A.D.
Corwin, J. He, and G.M. Zassenhaus$^{c}$}

\address{\hfill {\it Laboratory of Atomic and Solid State Physics and 
The Cornell Center for Materials Research,
Clark Hall, Cornell University, Ithaca, NY
14853-2501}}

\date{\today}
\maketitle

\begin{abstract}  Transition temperature data obtained as a function of particle
density in the $^4$He-Vycor system are compared with recent theoretical calculations for 3D
Bose condensed systems.  In the low density dilute Bose gas regime we find, in agreement with
theory, a positive shift in the transition temperature of the form $\Delta T/T_0 = \gamma
(na^{3})^{1/3}$.  At higher densities a maximum is found in the ratio of $T_c /T_0$ for
a value of the interaction parameter, na$^3$, that is in agreement with path-integral Monte
Carlo calculations.

%\pacs{PACS numbers: 03.75.Fi, 05.30.Jp, 67.40.-w}
\medskip

\end{abstract}

}

The role of interparticle interactions in the determination of the properties of low density
Bose-Einstein condensed (BEC) systems has been a topic of interest for many years.  In spite
of a long history of theoretical investigation \cite{Lee} dating back to the 1950¹s, elementary
questions such as the possible shift in the transition temperature, $T_{c}$, with density and
interaction strength have remained unsettled until the recent past.  In the case of the
repulsive interactions in the dilute Bose gas, there has now emerged a consensus \cite{Stoof} \cite{Gruter} \cite{Baym}  that
$T_{c}$ will be an increasing function of the interaction parameter, na$^3$, where a is the
hard sphere diameter and n the particle density.  This may seem a surprising result, since it
is well known that in the case of liquid $^4$He the superfluid transition occurs at a
temperature well below the transition temperature, $T_{0}$, for an ideal Bose gas with the
same particle mass and number density.  Moreover, a number of the earlier calculations \cite{Giraradeau} had found that the transition temperature would be reduced as a consequence of interparticle
interaction.

Motivated by the recent theoretical work in this area, we have examined the
dependence of the transition temperature on superfluid particle number for the $^{4}$He-Vycor
system.  In our early work with this system we demonstrated, for the first time in 1983 \cite{Crooker},
an experimental realization of the weakly interacting or ``dilute'' Bose gas.  The lowest
density achieved in these experiments was on the order of $2 \times 10^{18}$ per cm$^3$. 
This is sufficiently low to provide a region of overlap, in terms of the interaction
parameter, with the values of na$^{3}$, currently accessible to the BEC systems realized with
$^{87}$Rb \cite{Anderson} or
$^{23}$Na \cite{Davis} atoms confined within magnetic or optical traps.  In the case of Bose
condensed atomic hydrogen \cite{Fried} the small s-wave scattering length and limits on the particle
density, set by recombination, restrict this system to values of na$^3$ several orders of
magnitude below the values that can be achieved with Bose-condensed Na, Rb or He.  

For questions such as
the effect on
$T_{c}$ of increasing the interaction parameter, the $^4$He-Vycor system offers advantages over
the BEC systems of trapped atomic gases, because in the Vycor case, the interaction parameter
can be varied continuously from the low-density, weakly interacting limit to the strongly
interacting regime currently inaccessible with the alkali vapor systems.  Further, working with
the
$^4$He-Vycor system allows much larger sample sizes, on the order a cm$^3$, with an
essentially unlimited time for observation of the BEC superfluid state.

In this letter we will give only a brief summary of our experimental methods.  The cryogenic
techniques and thermometry methods are those standard in the study of superfluid $^3$He, and
the reader is referred elsewhere \cite{Richardson} for details.  For the present experiments, we have used a
torsional oscillator technique to obtain a signal proportional to the superfluid particle
density within our Vycor sample, which then allows us to estimate the number density for the
particles participating in the superfluid and as well as to determine the superfluid
transition temperature. The interior channels of the porous Vycor glass used for
these measurements range in diameter from 4 to 8 nm and form a highly interconnected
3D network.  The superfluid helium atoms are constrained by van der Waals forces to move over
the complex 3D-connected surface provided by the pores.  It is important to appreciate that at
low temperatures the thermal wavelength of the mobile superfluid particles can be larger than
the pore size and that the Feynman exchange cycles characterizing the BEC or superfluid state
will link many pores at low particle density \cite{Ceperley}.  Therefore we model the superfluid phase as
a homogeneous Bose gas constrained within the volume of the Vycor sample.  We anticipate,
however, that the influence of the substrate and pore geometry may be reflected in a small
modification of the effective mass of the mobile superfluid atoms.  We then expect to observe,
in the low density limit, thermodynamic properties similar to those of the ``free'' or ideal
Bose gas with an effective mass, m$^*$.

In the discussion of our experimental results, we shall first consider the data for the
dilute Bose gas regime.  The quantities required for a comparison with theory are the
particle number density and the corresponding transition temperature.  The number density is
estimated from an extrapolation of the superfluid signal to zero temperature and the
calibrated mass sensitivity of our torsional oscillator.  Following the recent calulations \cite{Stoof} \cite{Gruter} \cite{Baym}, we expect that for low densities T$_c$ will be given by 

\begin{equation} 
T_{c} = T_{0} [ 1 + \gamma
(na^{3})^{1/3}], 
\end{equation}
where $T_{0}  = [2 \pi \hbar^{2} /  m^{*} k_{B} \zeta (3/2)^{2/3}] n^{2/3}$ is the transition temperature
for an ideal Bose gas with particle mass m$^*$ and density n.  The coefficient for $\gamma$ is
positive as required for T$_c$  to rise above T$_0$ as the interaction parameter,
na$^3$ increases.  Although there is agreement on the form of Eq. 1, the theoretical estimates
for
$\gamma$ range over more than an order of magnitude, from an estimate of 0.34 reported by
Gr$\ddot u$ter, Ceperley, and Lalo$\ddot e$ (GCL) \cite{Gruter}, 2.9 by Baym et al. \cite{Baym}, to a value of
4.66 given by Stoof \cite{Stoof}.

For a convenient comparison to experimental results we cast Eq. 1 in a linear form with the
variable, n$^{1/3}$a, as 

\begin{equation}
\frac {T_{c}} {n^{2/3}} =\frac{2\pi \hbar^{2}} {  m^{*} k_{B} \zeta(3/2)^{2/3}} [1 +
\gamma(n^{1/3}a)].
\end{equation}
In Figure 1, we then plot T$_{c} / n^{2/3}$ against the parameter n$^{1/3}$a, taking a value of
0.22 nm \cite{Runge} for the helium hard core diameter.  As expected from theory, a
linear fit gives a good representation for our data in the low density regime.  The zero
density intercept of the fit serves to determine an effective mass ratio of m$^*$/m = 1.34,
where m is the $^4$He mass, and the slope yields a value of 5.1 for $\gamma$, which is within a
factor of 1.09 of the value given by Stoof.   

\begin{figure}[t!]
\parbox{2.6in}{
\includegraphics{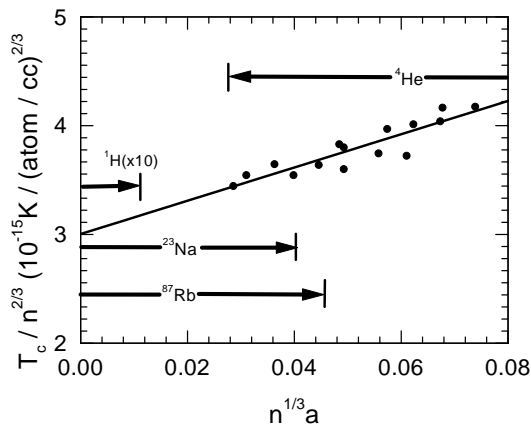}
\vspace{2.4in}

}
\caption{The quantity $T_{c} / n^{2/3}$ is plotted as a function of the parameter
n$^{1/3}$a.  The straight line is our fit to these data.  The arrows indicate the range of
interaction parameter covered in the $^1$H, $^4$He, $^{87}$Rb, and $^{23}$Na experiments. Note
that in the case of $^1$H the scale has been enlarged by a factor of 10.}
\end{figure}

It is of interest to compare the range of the interaction parameter explored in the various BEC
experiments.  The highest particle densities achieved to date are 4x10$^{14}$ 
atoms/cm$^3$ in the case of $^{87}$Rb and 3x10$^{15}$ cm$^3$ for $^{23}$Na, however, for
both the alkali atoms the s-wave scattering lengths are much larger than for the respective
value for the $^4$He atom.  The larger scattering lengths for the alkalies more than compensate for the lower
particle densities and result in a range of overlap in terms of
the interaction parameter between all three systems.  This is illustrated in  Figure 1, where
we indicate the range of the quantity n$^{1/3}$a for the $^4$He, experiments 
$^{23}$Na and $^{87}$Rb experiments.  Since the $^4$He-Vycor system clearly exhibits
superfluid properties, one may also expect superfluidity to exists in the Bose-condensed
alkali systems at comparable values of the interaction parameter and sufficiently low
temperature.

\begin{figure}[t!]
\parbox{2.6in}{
\includegraphics{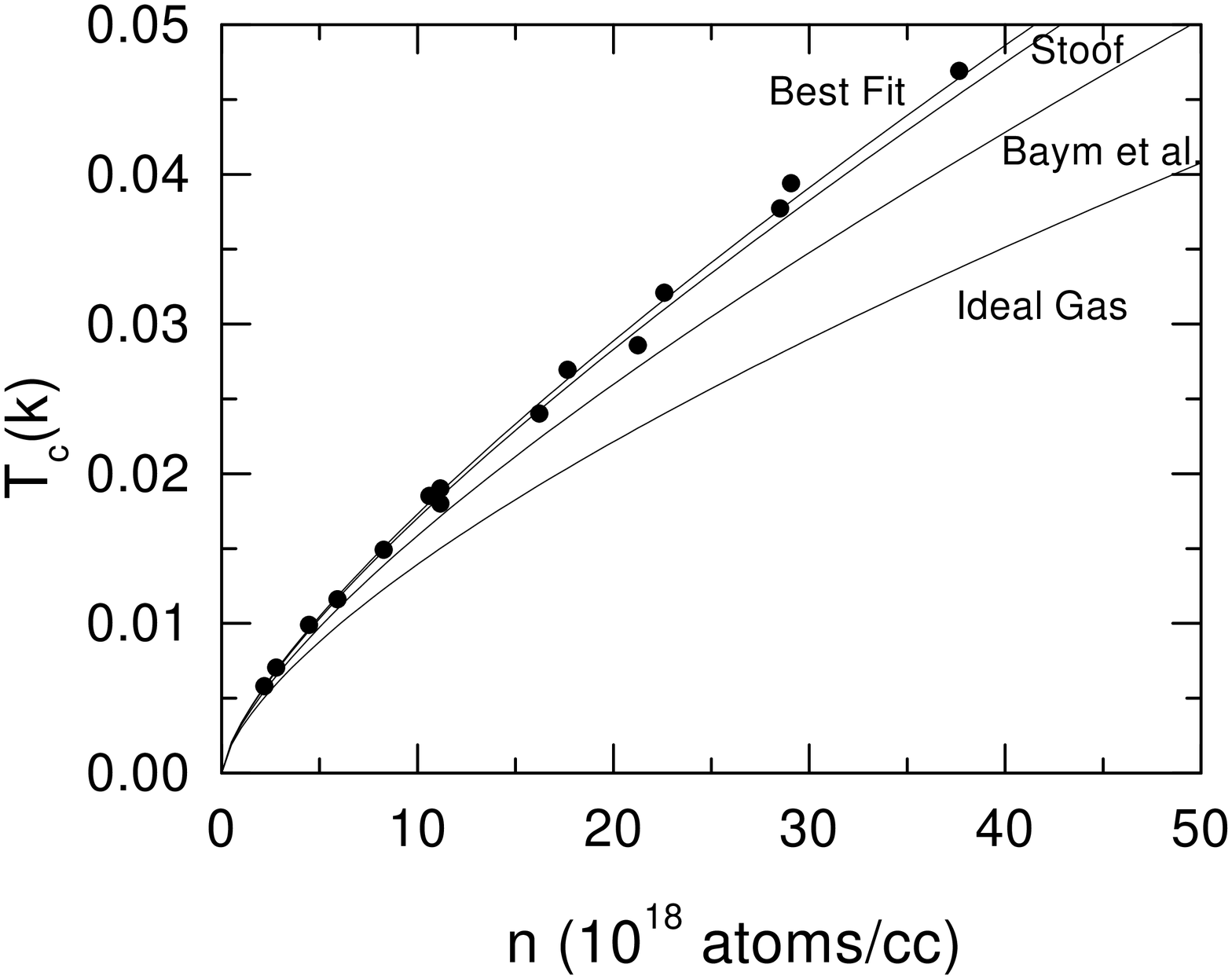}
\vspace{2.4in}

}
\caption{The observed transition temperature, T$_c$, is plotted as a function of the
particle density, n.  The top curve through the data is the function given by Eq. 1 with the
parameters  for the effective mass, m$^*$ and $\gamma$ determined as from the linear fit shown in Fig.
1.  The two other curves are for $\gamma$ = 4.66 as given by Stoof and $\gamma$ = o, the noninteracting
case.}
\end{figure}

In Figure 2 we show a more conventional plot of the low density data.  Here we have plotted
T$_c$ as a function of the particle density, n.  The curve through the data is the
theoretical expression given in Eq. 1 with our best fit values for $\gamma$ and m$^*$ as
determined previously.  Two curves based on the $\gamma$ estimates of Stoof and Baym et al. are
also shown.  The lowest curve is for the case of the noninteracting Bose gas case (i.e.,
$\gamma$ = 0 in Eq. 1).  The relatively close agreement with the estimate of Stoof is
clear in this plot.

In contrast to the other theoretical treatments of the BEC problem for the interacting
Bose gas \cite{Stoof} \cite{Baym}, the calculations of GCL are not restricted to the low density limit.  The
path-integral Monte Carlo technique employed by GCL allows a prediction for the ratio of
T$_c$/T$_0$ over the entire range of na$^3$, from the dilute gas limit to densities approaching
the freezing point of liquid helium.  In the top panel of Figure 3 we show the results of the
GCL calculation taken from Ref. \cite{Gruter}.  At low values of the interaction parameter, GCL also find
that the ratio, T$_c$/T$_0$, increases in proportion to n$^{1/3}$a in agreement with the
calculation of Stoof and of Baym et al.   An interesting feature of the GCL
calculation is the maximum in T$_c$/T$_0$ found near a value of the interaction parameter
na$^3$ = 0.01.

In the lower panel of Figure 3, we show the data obtained in a number of different
$^4$He-Vycor experiments at Cornell, including a recent measurement designed  specifically to
map out the region of the peak found by GCL.  Although the higher
density data show more scatter than the data of Figs. 1 and 2, the trends are clear, and
there is a pleasing agreement in the qualitative form of the experimental data as compared to
the GCL calculation.  In particular, there is an excellent match in the position of the peak
value for the ratio T$_c$/T$_0$ as a function of the interaction parameter.  A remaining
problem, however, is the as yet unexplained disagreement in the magnitude of the effect 
estimated by GCL for low densities as compared to our experimental data or the
calculations of Stoof \cite{Stoof} and Baym et al. \cite{Baym}.

\begin{figure}[t!]
\parbox{2.8in}{
\includegraphics{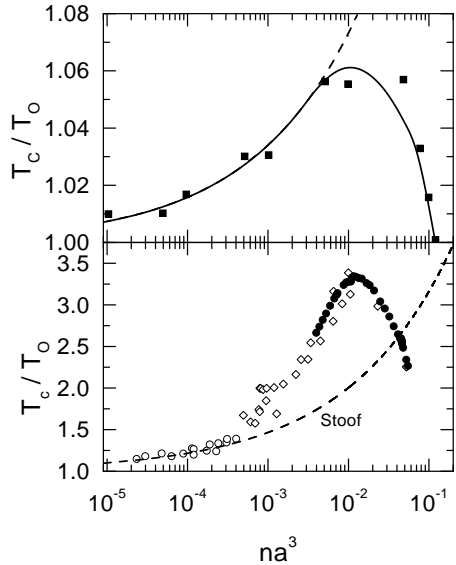}
\vspace{3.2in}

}
\caption{The top panel shows the values of the ratio $T_{c} /T_{0}$ calculated by GCL
as a function of the interaction parameter na$^3$.  The curve through the lower density points
is given by $T_{c} /T_{0}  = 1 + 0.34 (na^{3})^{1/3}$.  The curve through the higher density
data including the region of the peak is merely a guide for the eye.  In the lower panel we
plot the values of $T_{c} /T_{0}$, obtained from a number of different Vycor experiments
(distinguished by different symbols), as a function of na$^3$.  The dashed curve is the
theoretical estimate of Stoof [2].}
\end{figure}

In conclusion, we wish to emphasis that studies with low density  Bose condensed helium
systems can provide a useful and illuminating contrast to the trapped gas systems for the
study of properties of the weakly interacting BEC fluids.  Particularly important aspects of the
$^4$He-Vycor system are the ability to study long term phenomena, such as the question of
robustness or stability of flow states in a Bose-condensed superfluid, and to extend
observations of BEC to the realm of larger interaction parameters than are accessible to the
trapped gas systems.

%\bigskip
%\centerline{\bf Acknowledgements}

\acknowledgments

The authors wish to acknowledge stimulating conversations with Gordon Baym, David Ceperley,
Michael Fisher, Geoffrey Chester, Veit Elser, Wolfgang Ketterle, and many others.  JDR would
like to thank the Aspen Center for Physics for its hospitality during the time the present
letter was conceived.  GMZ and ADC have been supported by the US Dept. of Ed. through the
GAANN Program-Physics Grants P200A80709 and P200A70615-98.  This work is supported by the
National Science Foundation through Grants No. DMR-9971124 and DMR-9705295, and by the Cornell
Center for Materials Research, through Grant No. DMR-9632275, CCMR Report No. 8405.

\bigskip

\noindent a.  Present address:  Physics Dept., Fordham University, Bronx, NY 10458-5198.

\noindent b.  Present address:  CNRS-CRTBT, BP 166, Grenoble Cedex 38042 France.

\noindent c.  Present address:  McKinsey \& Company, Three Landmark Square, 1st Floor,
Stamford, CT  06901.

%\medskip

\end{document}